\documentclass[prd,preprint,tightenlines,showpacs,floatfix,showpacs,preprintnumbers,nofootinbib,eqsecnum,superscriptaddress]{revtex4}

 \usepackage[dvips,final]{graphicx}
  \usepackage{amssymb}
   \usepackage{amsmath}
    \usepackage{amsfonts}
     \usepackage{epsfig}
      \usepackage{bm}% bold math
\begin{document}

\title{
Kinematical correlations of dielectrons \\
from semileptonic decays of heavy mesons\\
and Drell-Yan processes at BNL RHIC
}

\author{R. Maciu{\l}a}
\email{rafal.maciula@ifj.edu.pl}
\affiliation{Institute of Nuclear Physics PAN, PL-31-342 Cracow,
Poland} 

\author{A. Szczurek}
\email{Antoni.Szczurek@ifj.edu.pl}
\affiliation{Institute of Nuclear Physics PAN, PL-31-342 Cracow,
Poland} 
\affiliation{University of Rzesz\'ow, PL-35-959 Rzesz\'ow, Poland}

\author{G. {\'S}lipek}
\email{gabriela.slipek@ifj.edu.pl}
\affiliation{Institute of Nuclear Physics PAN, PL-31-342 Cracow,
Poland} 

\date{today}

\begin{abstract}
We discuss kinematical correlations between charged leptons from
semileptonic decays of open charm/bottom, leptons produced in
the Drell-Yan mechanism as well as some other mechanisms not
included so far in the literature in proton-proton scattering at BNL RHIC.
The distributions of charm and bottom quarks/antiquarks are calculated
in the framework of the $k_t$-factorization approach.
For this calculation we use different unintegrated parton distributions from the literature. 
The hadronization of heavy quarks is done with the help of well-known fragmentation functions.
Uncertainties of our predictions related to heavy quark masses, factorization and renormalization
scales as well as due to the choice of fragmentation model are also discussed. We use semileptonic decay
functions found by fitting recent semileptonic data obtained by 
the CLEO and BABAR collaborations. 
The Drell-Yan processes were calculated including transverse momenta of
quarks and antiquarks, using the Kwieci\'nski parton distributions.
We have also took into consideration reactions initiated by purely 
QED $\gamma^*\gamma^*$-fusion in elastic and inelastic pp collisions 
as well as recently proposed diffractive mechanism of exclusive 
charm-anticharm production.
The contribution of the later mechanism is rather small.
We get good description of the dilepton invariant mass spectrum 
measured recently by the PHENIX collaboration and present predictions
for the dilepton pair transverse momentum distribution
as well as distribution in azimuthal angle between electron and positron.
\end{abstract}


\maketitle

%---------------------------
\section{Introduction}
%---------------------------

Recently the PHENIX collaboration has measured dilepton invariant mass
spectrum from $0$ to $8$ GeV in proton-proton collisions 
at $\sqrt{s}=200$ GeV \cite{PHENIX}.
It is commonly believed that the main contribution
to the dielectron continuum comes from so-called nonphotonic electrons
which are produced mainly in semileptonic decays of charm and
bottom mesons. Up to now, production of open charm and bottom 
was studied only in inclusive measurements of charmed mesons 
\cite{Tevatron_mesons} and electrons \cite{electrons} and only inclusive
observables were calculated in pQCD approach \cite{cacciari,LMS09}. 
Such predictions give rather good description of the experimental data,
however, the theoretical uncertainties are quite large which makes the
situation somewhat clouded and prevents definite conclusions.

Some time ago we have studied kinematical correlations of $c \bar c$
quarks \cite{LS06}, which is, however, difficult to study experimentally.
High luminosity and in a consequence better statistics at present
colliders gives a new possibility to study not only inclusive
distributions but also correlations between outgoing particles
(meson-meson, meson-electron or electron-electron). Kinematical
correlations constitute an alternative method to pin down 
the cross section for charm and bottom production. 
It gives also a great possibility to separate charm and bottom 
contributions which has a crucial meaning for understanding 
the character of heavy quarks interactions with the matter 
created in high energy nuclear collisions \cite{mischke}.

%-----------------------
\section{Formalism}
%-----------------------

The $k_t$-factorization method is very useful to study correlations
between $c \bar c$ \cite{LS06} and $e^+ e^-$ from the Drell-Yan
processes \cite{SS08}.
In our calculations we take under consideration not only 
leptons from open charm/bottom decays but also leptons produced in 
Drell-Yan proccess, as well as leptons coming from elastic and 
inelastic processes initiated by photon-photon fusion. In the case of 
elastic reaction we follow exact momentum space calculations 
with 4-body phase space (see e.g. \cite{LS2010}) 
and for inelastic scattering we have applied unique
(collinear) photon distributions in the nucleon MRST2004 \cite{MRST}.

%----------------------------------------------------------------
\subsection{Dileptons from semileptonic decays}
%----------------------------------------------------------------

\label{sec:semileptonic_decays}

\begin{figure}[!h]
\begin{minipage}{0.45\textwidth}
 \centerline{\includegraphics[width=1.0\textwidth]{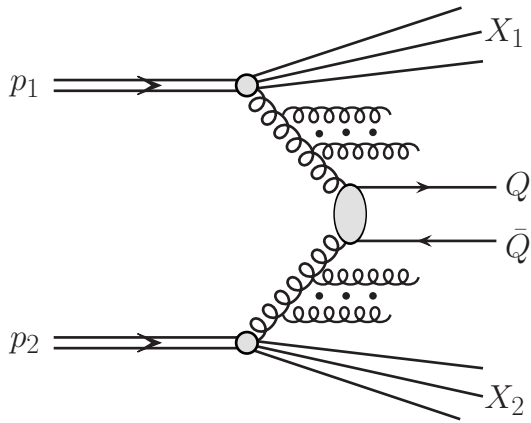}}
\end{minipage}
   \caption{The dominant mechanism of the $c \bar c$ and $b \bar b$
production at high energy. The emission of several extra gluons
is included in the unintegrated gluon (parton) distributions
used in the formalism.
 \small 
}
 \label{fig:inclusive}
\end{figure}

The electrons from semileptonic decays are produced in a three-stage process.
The whole procedure can be written in the following schematic way:
\begin{equation}
\frac{d \sigma^e}{d y d^2 p} =
\frac{d \sigma^Q}{d y d^2 p} \otimes
D_{Q \to h} \otimes
f_{h \to e} \; ,
\label{whole_procedure}
\end{equation}
where the symbol $\otimes$ denotes a generic convolution.
The first term is responsible for production
of heavy quarks/antiquarks (see Fig.\ref{fig:inclusive}). Next step is the process of formation of
heavy mesons and the last ingredient 
is semileptonic decay of heavy mesons to electrons/positrons.
The inclusive production of heavy quark/antiquark pairs can be calculated
in the framework of the $k_t$-factorization \cite{CCH91}.
In this approach transverse momenta of initial partons are included and
emission of gluons is encoded in a so-called unintegrated gluon,
in general parton, distributions.
In the leading-order approximation within the $k_t$-factorization approach
the differential cross section for the $Q \bar Q$ or Drell-Yan process
can be written as:
\begin{eqnarray}
\frac{d \sigma}{d y_1 d p_{1t} d y_{2} d p_{2t} d \phi} =
\sum_{i,j} \; \int \frac{d^2 \kappa_{1,t}}{\pi} \frac{d^2 \kappa_{2,t}}{\pi}
\frac{1}{16 \pi^2 (x_1 x_2 s)^2} \; \overline{ | {\cal M}_{ij} |^2}\\
\nonumber 
\delta^{2} \left( \vec{\kappa}_{1,t} + \vec{\kappa}_{2,t} 
                 - \vec{p}_{1,t} - \vec{p}_{2,t} \right) \;
{\cal F}_i(x_1,\kappa_{1,t}^2) \; {\cal F}_j(x_2,\kappa_{2,t}^2) \; , \nonumber \,\,
\end{eqnarray}
where ${\cal F}_i(x_1,\kappa_{1,t}^2)$ and ${\cal F}_j(x_2,\kappa_{2,t}^2)$
are the unintegrated gluon (parton) distribution functions (UPDFs). 
The longitudinal momentum fractions can be calculated as
\begin{eqnarray}
x_1 &=& \frac{m_{1t}}{\sqrt{s}} \exp( y_1) + \frac{m_{2t}}{\sqrt{s}}
\exp( y_2) \; ,                                       \nonumber \\ 
x_2 &=& \frac{m_{1t}}{\sqrt{s}} \exp(-y_1) + \frac{m_{2t}}{\sqrt{s}}
\exp(-y_2)  \; ,
\end{eqnarray}
where $y_1$ and $y_2$ are rapidities of heavy quark and heavy antiquark,
and $m_{1t}$ and $m_{2t}$ are their transverse masses.

%--------------------------------------------------------------------------
%-----------------------------------------------------------------------------

There are two types of the LO $2 \to 2$ subprocesses which contribute
to heavy quarks production, $gg \to Q \bar Q$ and $q \bar q \to Q \bar
Q$.
The first mechanism dominates at large energies and the second one
near the threshold. Only $g g \to Q \bar Q$ mechanism is included here. We use off-shell matrix elements
corresponding to off-shell kinematics so hard amplitude depends on transverse momenta
(virtualities of initial gluons) in the exact way.
At relatively low RHIC energies rather intermediate $x$-values become relevant so the Kwiecinski UGDFs
seem applicable in this case \cite{Kwiecinski}. However, to show the uncertainty of our predictions resulting from
different approaches in calculating uninegrated parton distributions we have also used
Kimber-Martin-Ryskin (KMR) \cite{KMRupdf} and Kutak-Stasto models \cite{Kutak}. All of them have different theoretical background.
It is therefore very interesting to compare such results with the PHENIX data 
and verify applicability of these UGDFs at RHIC.
In the case of the Kwiecinski distributions we fix the renormalization and factorization scales to standard
values $\mu_R^2 = \mu_F^2 = 4 m_Q^2$. Using the KMR UGDF the mostly used set of these parameters
in the context of inclusive heavy quark production is $\mu_R^2 = 4 m_Q^2$ and $\mu_F^2 = M_{Q\bar{Q}}^2$,
where $M_{Q\bar{Q}}$ is the invariant mass of the $Q\bar{Q}$ pair.

The hadronization of heavy quarks is usually done
with the help of fragmentation functions. The inclusive distributions of
hadrons can be obtained through a convolution of inclusive distributions
of heavy quarks/antiquarks and Q $\to$ h fragmentation functions:
\begin{equation}
\frac{d \sigma (y_1, p_{1t}^{H}, y_{2}, p_{2t}^{H}, \phi)}{d y_1 d p_{1t}^{H} d y_{2} d p_{2t}^{H} d \phi}
 \approx
\int \frac{D_{Q \to H}(z_{1})}{z_{1}}\cdot \frac{D_{\bar Q \to \bar H}(z_{2})}{z_{2}}\cdot
\frac{d \sigma (y_1, p_{1t}^{Q}, y_{2}, p_{2t}^{Q}, \phi)}{d y_1 d
  p_{1t}^{Q} d y_{2} d p_{2t}^{Q} d \phi} d z_{1} d z_{2} \; ,
\end{equation}
where: 
{$p_{1t}^{Q} = \frac{p_{1t}^{H}}{z_{1}}$, $p_{2t}^{Q} =
  \frac{p_{2t}^{H}}{z_{2}}$, where
meson longitudinal fractions  $z_{1}, z_{2}\in (0,1)$.
We have made approximation assuming that $y_{1}, y_{2}, \phi$  are
unchanged in the fragmentation process.

There are several models of fragmentation functions in the literature. Here we mostly use the 
Peterson fragmentation function \cite{Peterson}. However, to check the
sensitivity of our results to the choice of the fragmentation model, we have also applied fragmentation
functions proposed by Kartvelishvili et al. \cite{kartvel} and Braaten et al. \cite{bcfy}.

Recently the CLEO and BABAR collaborations have measured very precisely
the spectrum of electrons/positrons coming from
the weak decays of $D$ and $B$ mesons, respectively \cite{CLEO}.
These functions can in principle be calculated.
This introduces, however, some model 
uncertainties and requires inclusion of all final 
state channels explicitly. An alternative is to use 
proper experimental input which after renormalizing 
to experimental branching fractions
can be use to generate electrons/positrons 
in a Monte Carlo approach. The electrons (positrons) are generated isotropically
in the heavy meson rest frame.
In the present paper we use parametrizations of the decay functions
found in Ref.\cite{LMS09}.

%-----------------------------------------
\subsection{Drell-Yan dileptons}
%-----------------------------------------

\label{sec:Drell_Yan}

The electron and positron produced in the Drell-Yan mechanism are
naturally correlated. We have shown recently \cite{SS08} how to use
the transverse momentum dependent parton (quark, antiquark)
distributions to obtain several differential distributions.
A basic diagram of the mechanism is shown in Fig.\ref{fig:drell-yan}.
In our calculations here we follow Ref.\cite{SS08} and use the Kwiecinski parton distributions.
Here the off-shellness of quark/antiquark is included in the kinematics
and the matrix element taken here in the on-shell form expressed in terms of the subprocess invariants
calculated with the off-shell condition.
 This is not fully consistent but avoids problems when on-shell momenta of quarks
 and on-shell matrix element are used \cite{Wong,Murgia}.
In any case the result of our approach is not very different than that
for the collinear approach but includes kinematical effect of transverse momenta
which is crucial to understand e.g. azimuthal correlations and
distributions in transverse momentum of the dilepton pair, impossible
to address in the collinear approach.

The differential cross section for the 0-th order contribution
including quark/antiquark transverse momenta can be written as:
\begin{equation}
\begin{split}
\frac{d \sigma}{d y_1 d y_2 d^2p_{1t} d^2p_{2t}} = \sum_{f} \;
\int \frac{d^2 \kappa_{1t}}{\pi} \frac{d^2 \kappa_{2t}}{\pi}
\frac{1}{16 \pi^2 (x_1 x_2 s)^2} \; \\
%\nonumber 
\delta^2 \left( \vec{\kappa}_{1t} + \vec{\kappa}_{2t}
                 - \vec{p}_{1t} - \vec{p}_{2t} \right) \; 
[{\cal F}_{q_f}(x_1,\kappa_{1t}^2,\mu_F^2) \; {\cal F}_{\bar q_f}(x_2,\kappa_{2t}^2,\mu_F^2)\;
\overline{|M({q \bar q} \to {e^+ e^-})|^2 } \; \\
+ {\cal F}_{\bar q_f}(x_1,\kappa_{1t}^2,\mu_F^2) \; {\cal F}_{q_f}(x_2,\kappa_{2t}^2,\mu_F^2) \;
\overline{|M({q \bar q} \to {e^+ e^-})|^2 } \; ] \; ,
\end{split}
\label{0th_order_kt-factorization}
\end{equation}
where
${\cal F}_i(x_1,\kappa_{1t}^2)$ and ${\cal F}_i(x_2,\kappa_{2t}^2)$
are unintegrated quark/antiquark distributions in hadron $h_1$ and $h_2$,
respectively.

The longitudinal momentum fractions are evaluated in terms of
final lepton rapidities and transverse momenta:
\begin{eqnarray}
x_1 &=& \frac{m_{1t}}{\sqrt{s}}\exp( y_1) +\frac{m_{2t}}{\sqrt{s}}\exp( y_2) \;,\nonumber \\
x_2 &=& \frac{m_{1t}}{\sqrt{s}}\exp(-y_1) +\frac{m_{2t}}{\sqrt{s}}\exp(-y_2) ,
\end{eqnarray}
where $m_t = \sqrt{{p_t}^2 + m^2}$ are transverse masses of
electron and positron.

The delta function in Eq.(\ref{0th_order_kt-factorization}) can be 
eliminated as e.g. in Refs.\cite{LS06}.

%--------------------------------------------------------------------------
\begin{figure}[!h]
\begin{minipage}{0.4\textwidth}
 \centerline{\includegraphics[width=1.0\textwidth]{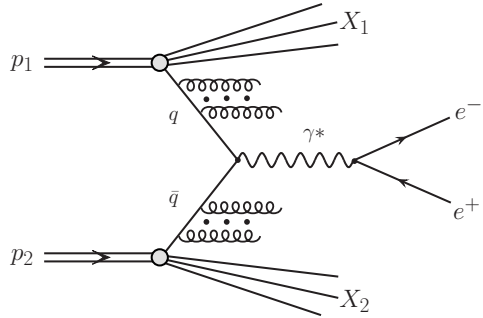} }
\end{minipage}
   \caption{
 \small The Drell-Yan mechanism of the dielectron
pair production. The extra gluon emissions are included in the formalism
of unintegrated parton distributions.
}
 \label{fig:drell-yan}
\end{figure}
%-----------------------------------------------------------------------------

%------------------------------------------------------------------------------
\subsection{QED \bm{$\gamma^*\gamma^{*}$} elastic and
  inelastic production of dileptons}
%------------------------------------------------------------------------------

\label{sec:qed}

The matrix element for the $pp \to ppe^{+}e^{-}$ reaction via 
$\gamma^*\gamma^*$-fusion (see Fig.\ref{fig:qed}) can be approximately written as
\begin{eqnarray}
{\cal M}^{\gamma^* \gamma^*} \approx e F_1(t_1) \frac{(p_1 +
p_1')^{\nu}}{t_1}\, V_{\mu\nu}^{\gamma^* \gamma^*}(q_1,q_2)\,
\frac{(p_2 + p_2')^{\mu}}{t_2}\, e F_1(t_2)\,, \label{ggamp}
\end{eqnarray}
where $F_1(t_1)$ and $F_1(t_2)$ are Dirac proton electromagnetic
form factors, and the $\gamma^*\gamma^* \to e^{+}e^{-}$ vertex
has the form
\begin{eqnarray}\nonumber
&&{}V_{\lambda_q\lambda_{\bar{q}},\mu\nu}^{\gamma^*
\gamma^*}=e^2\,\bar{u}_{\lambda_q}(k_1)
\biggl(\gamma^{\nu}\frac{\hat{q}_{1}-\hat{k}_{1}-m}
{(q_1-k_1)^2-m^2}\gamma^{\mu}-\gamma^{\mu}\frac{\hat{q}_{1}-
\hat{k}_{2}+m}{(q_1-k_2)^2-m^2}\gamma^{\nu}\biggr)v_{\lambda_{\bar{q}}}(k_2).\\
\label{ggvert}
\end{eqnarray}
The above formulae are used to calculate differential cross section
via exact integration in the full 4-body phase-space. The details
can be found e.g. in Ref. \cite{LS2010}. We shall call this contribution
double elastic for brevity.

In addition, there are components when one of the protons, or even both
(see Fig.\ref{fig:qed}), do not survive the collision.
Corresponding contributions will be called single and double inelastic
processes respectively. The double inelastic contribution can be calculated
as the gluon-gluon contribution in the parton model by replacing
gluon distributions in the nucleon by corresponding photon distributions.
Only one group discussed photon distributions in the nucleon
\cite{MRST}. The corresponding cross section can be calculated as
\begin{equation}
\frac{d \sigma}{d y_1 d y_2 d^2 p_t} = \frac{1}{16 \pi^2 {\hat s}^2} 
x_1 f_{\gamma/p}(x_1,\mu^2) x_2 f_{\gamma/p}(x_2,\mu^2) 
\overline{|{\cal M}_{\gamma \gamma \to e^+ e^-}|^2}  \; .  
\label{double_inelastic}
\end{equation}

%--------------------------------------------------------------------------
\begin{figure}[!h]
\begin{minipage}{0.22\textwidth}
 \centerline{\includegraphics[width=1.0\textwidth]{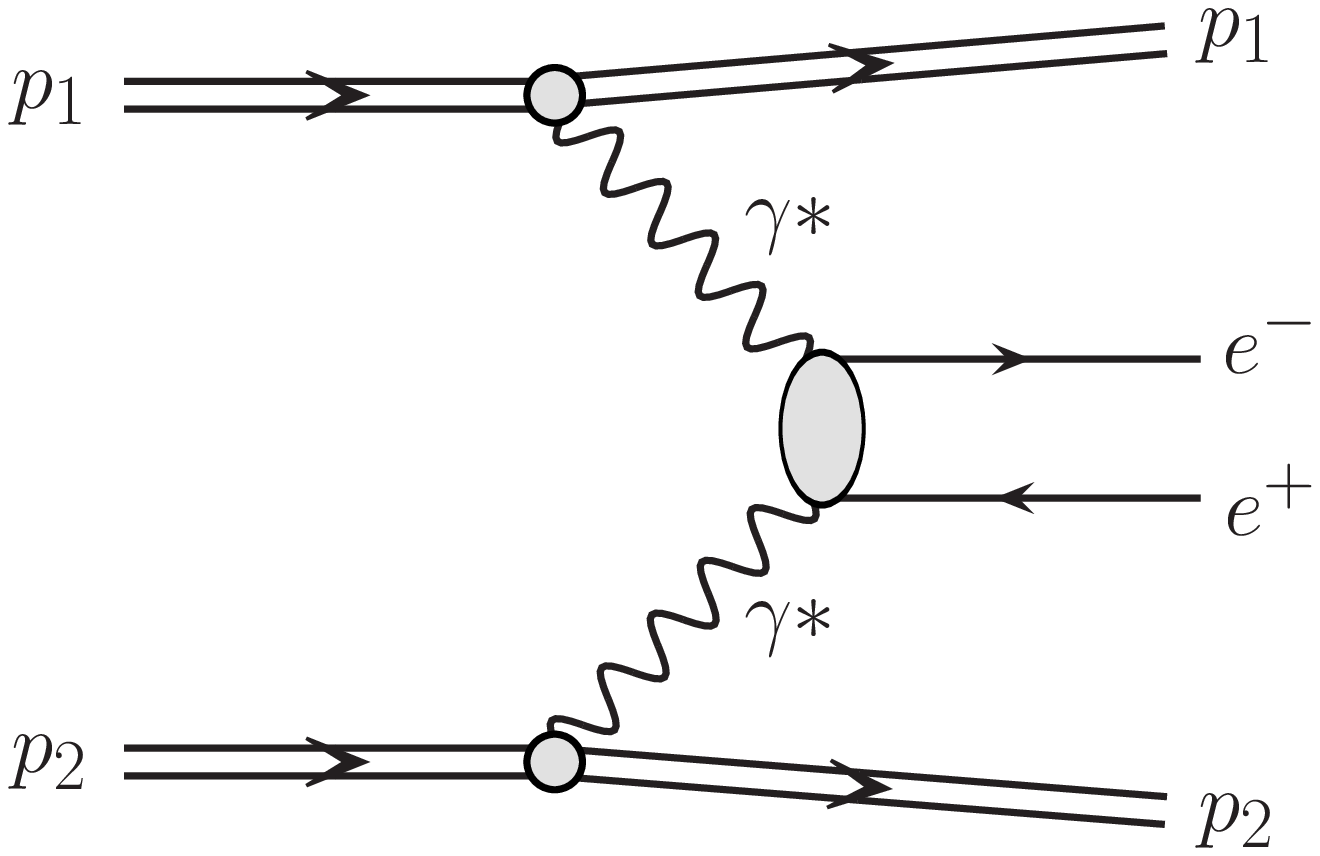}}
\end{minipage}
\hspace{0.5cm}
\begin{minipage}{0.22\textwidth}
 \centerline{\includegraphics[width=1.0\textwidth]{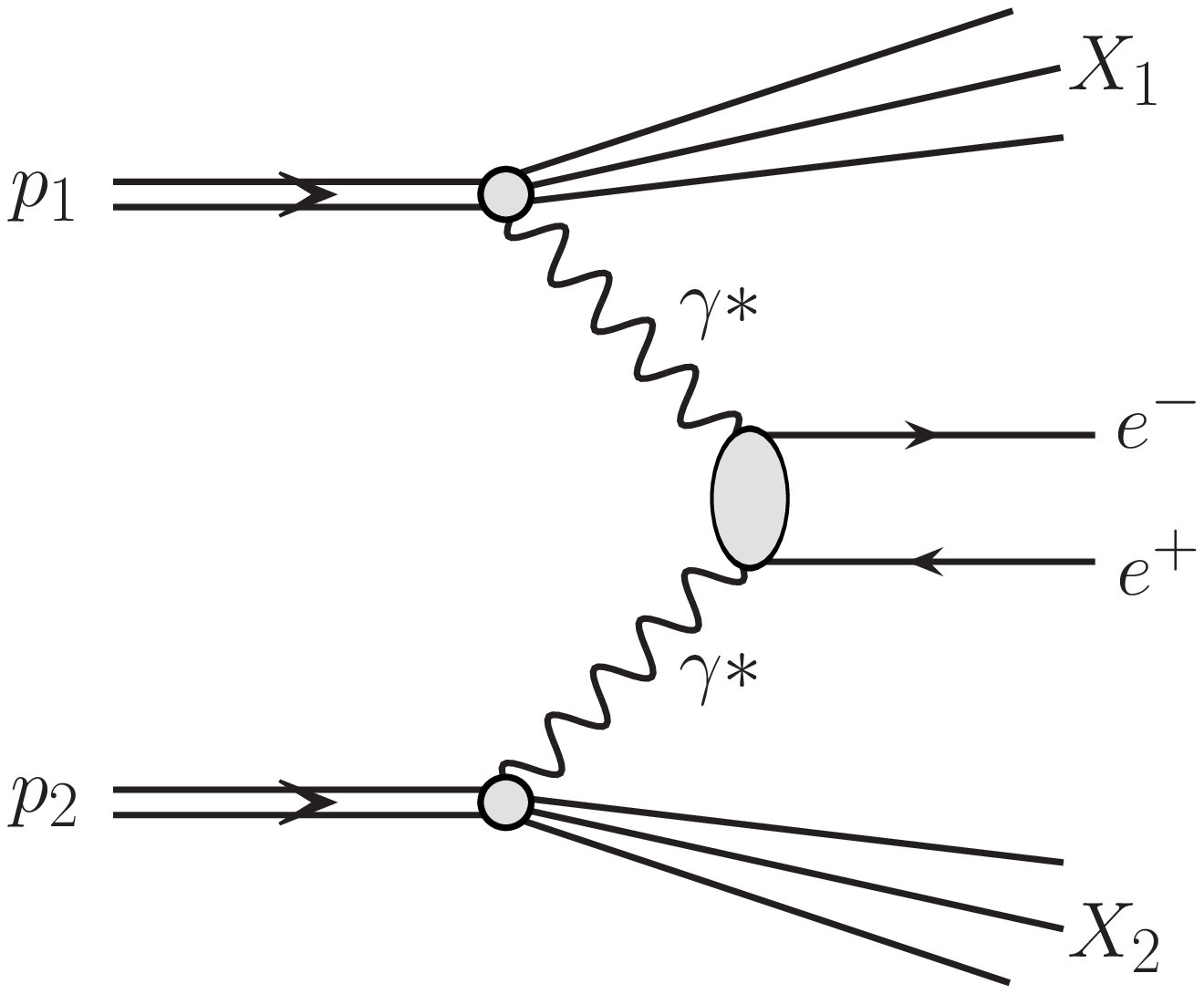}}
\end{minipage}
\begin{minipage}{0.22\textwidth}
 \centerline{\includegraphics[width=1.0\textwidth]{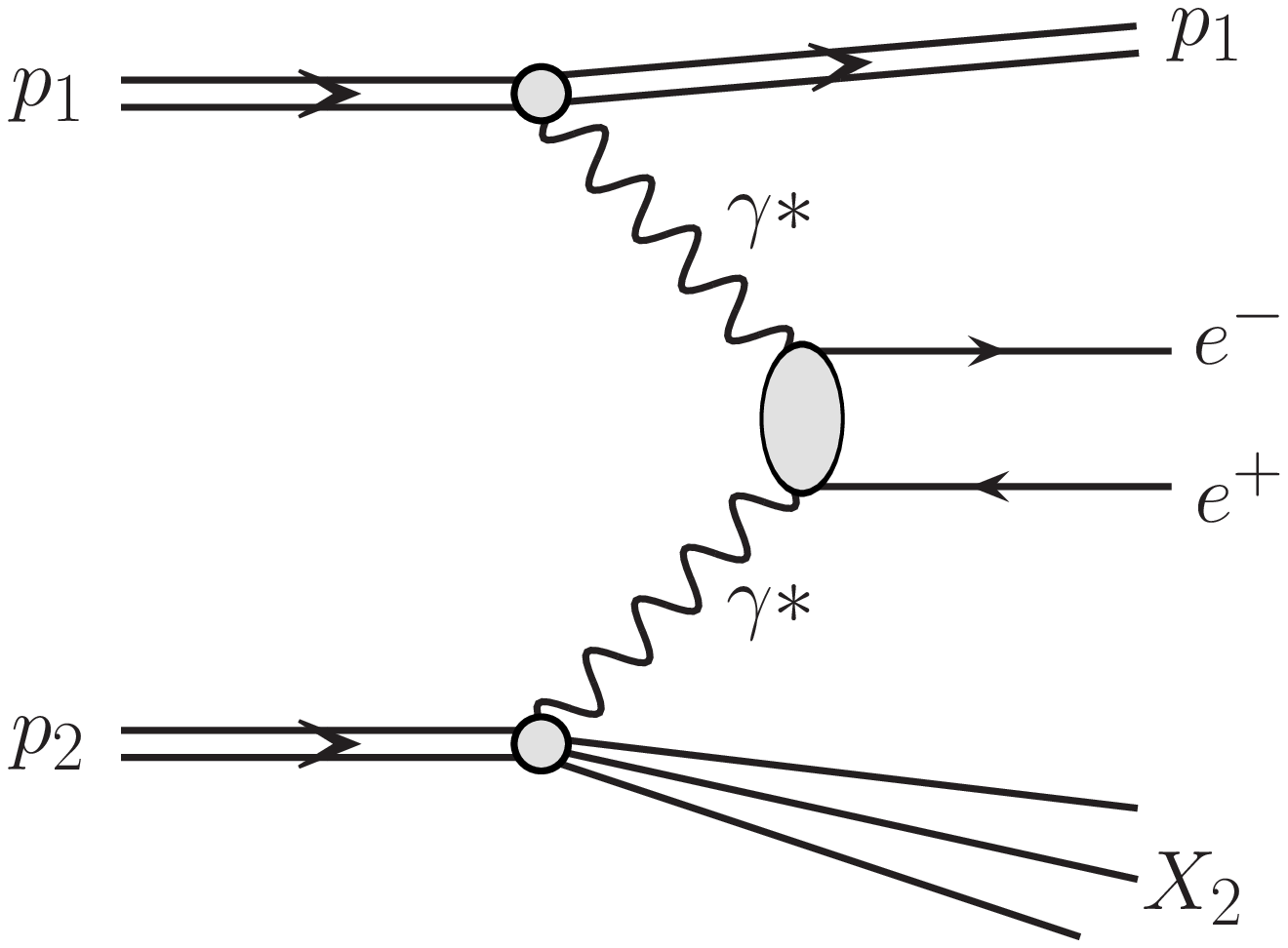}}
\end{minipage}
\hspace{0.5cm}
\begin{minipage}{0.22\textwidth}
 \centerline{\includegraphics[width=1.0\textwidth]{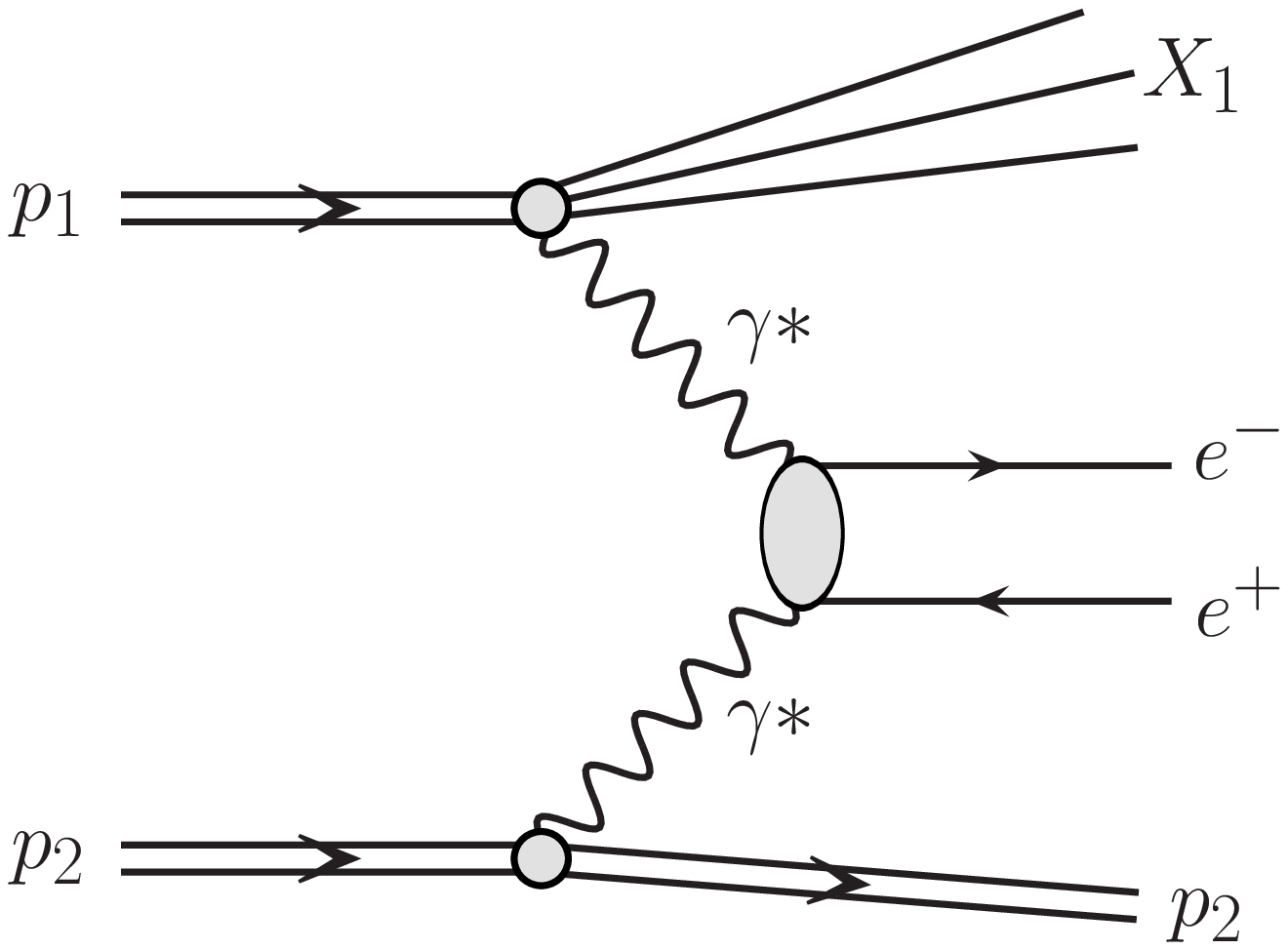}}
\end{minipage}
   \caption{
 \small Diagrammatic representation of processes initiate by
photon-photon subprocesses: double-elastic, 
double-inelastic, inelastic-elastic
and elastic-inelastic.
}
 \label{fig:qed}
\end{figure}
%-----------------------------------------------------------------------------

%
The matrix element can be found in several text books.
The cross section for single inelastic process can be calculated
by a replacement of one of $f_{\gamma/p}(x,\mu^2)$ by $f_{\gamma/p}^{el}(x)$
which is often called elastic photon flux factor. Relevant formulae can be found
in \cite{DZ}.

%-------------------------------------------------------------------------
\subsection{Exclusive double-diffractive production of open charm}
%-------------------------------------------------------------------------

\label{sec:edd}

Recently, our group has calculated, for the first time in the literature,
the exclusive double diffractive (EDD) production of open charm \cite{MPS2010}. 
A sketch of this mechanism is shown in Fig.~\ref{fig:EDD_mechanism}.

%-------------------------------------------------------------
\begin{figure}[h!]
\begin{minipage}{0.35\textwidth}
 \centerline{\includegraphics[width=1.0\textwidth]{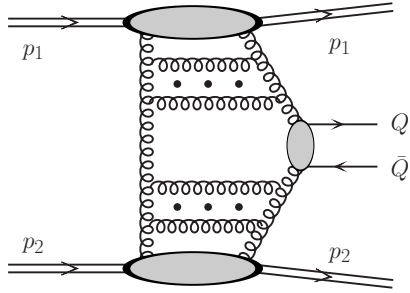}}
\end{minipage}
\caption{\small The mechanism of exclusive double-diffractive
production of open charm.}
\label{fig:EDD_mechanism}
\end{figure}
%-------------------------------------------------------------

The $p p \to p (c \bar c) p $ reaction is treated as a genuine 4-body
process with exact kinematics. This can be easily used to apply
kinematical cuts required by experiments. According to the Kaidalov-Khoze-Martin-Ryskin (KKMR)
approach used previously for the exclusive Higgs boson production
\cite{KMR_Higgs}, the amplitude of the exclusive 
diffractive $q\bar{q}$ pair production $pp\to p(q\bar{q})p$ can be
written as \cite{MPS2010}
\begin{eqnarray}
{\cal
M}_{\lambda_q\lambda_{\bar{q}}}^{p p \to p p q \bar q}(p'_1,p'_2,k_1,k_2) &=&
s\cdot\pi^2\frac12\frac{\delta_{c_1c_2}}{N_c^2-1}\,
\Im\int d^2
q_{0,t} \; V_{\lambda_q\lambda_{\bar{q}}}^{c_1c_2}(q_1, q_2, k_1, k_2) \nonumber \\
&&\frac{f^{\mathrm{off}}_{g,1}(x_1,x_1',q_{0,t}^2,
q_{1,t}^2,t_1)f^{\mathrm{off}}_{g,2}(x_2,x_2',q_{0,t}^2,q_{2,t}^2,t_2)}
{q_{0,t}^2\,q_{1,t}^2\, q_{2,t}^2} \; ,
\label{amplitude}
\end{eqnarray}
where $\lambda_q,\,\lambda_{\bar{q}}$ are helicities of heavy $q$
and $\bar{q}$, respectively. Above $f_{g,1}^{\mathrm{off}}$ and
$f_{g,2}^{\mathrm{off}}$ are the off-diagonal unintegrated gluon
distributions in nucleon 1 and 2, respectively.

The vertex factor
$V_{\lambda_q\lambda_{\bar{q}}}^{c_1c_2}=
 V_{\lambda_q\lambda_{\bar{q}}}^{c_1c_2}(q_1,q_2,k_1,k_2)$
in expression (\ref{amplitude}) is the production amplitude
of a pair of massive quark $q$ and antiquark $\bar{q}$ with
helicities $\lambda_q$, $\lambda_{\bar{q}}$ and
momenta $k_1$, $k_2$, respectively. The bare amplitude above is subjected to absorption corrections
which depend on collision energy and on the spin-parity
of the produced central system.

In the KMR approach the off-diagonal parton distributions
are calculated as
\begin{eqnarray}
f_g^{\mathrm{KMR}}(x,Q_{t}^2,\mu^2,t) &=& R_g
\frac{d[xg(x,k_t^2)S_{1/2}(k_{t}^2,\mu^2)]}{d \log k_t^2} |_{k_t^2
= Q_{t}^2} \;
F(t) \; ,
\label{KMR_off-diagonal-UGDFs}
\end{eqnarray}
where $S_{1/2}(q_t^2, \mu^2)$ is a Sudakov-like form factor relevant
for the case under consideration \cite{MR}. The factor $R_g$ here is the skewedness parameter 
and at the RHIC energy the value $ R_g \sim 1.4$ seems to be relevant.

In the present calculation we use standard GRV95 collinear gluon
distributions \cite{GRV95}. For this process we take the
renormalization and factorization scales to be $\mu_{R}^2=\mu_{F}^2=M_{c \bar
c}^2/4$ as in Ref.\cite{MPS2010}. Absorption effects are included approximately by
multiplying the cross section by the gap survival factor $S_G = 0.15$.
More details about EDD production of charm quarks can be found in our original paper \cite{MPS2010}.
In order to compare this predictions with the PHENIX data we have
applied the same procedure of hadronization and semileptonic decays as
in the case of inclusive processes described briefly in subsection \ref{sec:semileptonic_decays}.

%\newpage
%-----------------------------------
\section{Numerical results}
%-----------------------------------

Let us come now to the presentation of our results.
%In all results shown above the azimuthal angle acceptance of PHENIX
%detector \cite{PHENIX} is included.
In Fig.\ref{fig:sources-kwiec} we show distribution in dielectron
invariant mass. We have included several mechanisms.
The solid lines represent the contribution of charm (upper one)
and bottom (lower one) production calculated using the Kwiecinski UGDFs and subsequent 
semileptonic decays which was calculated 
in the way described in subsection \ref{sec:semileptonic_decays}.
The Drell-Yan contribution is shown by the long-dashed line
and its contribution is comparable to the contribution of semileptonic decays.
The gamma-gamma contributions (sum of the four contributions of diagrams 
in Fig.\ref{fig:qed}) is shown by the blue dashed line 
at the bottom-left corner of the figure. 
The very small EDD contribution is shown for completness by the green dotted line.
%--------------------------------------------------------------------------
\begin{figure}[!h]
\begin{minipage}{0.75\textwidth}
 \centerline{\includegraphics[width=1.0\textwidth]{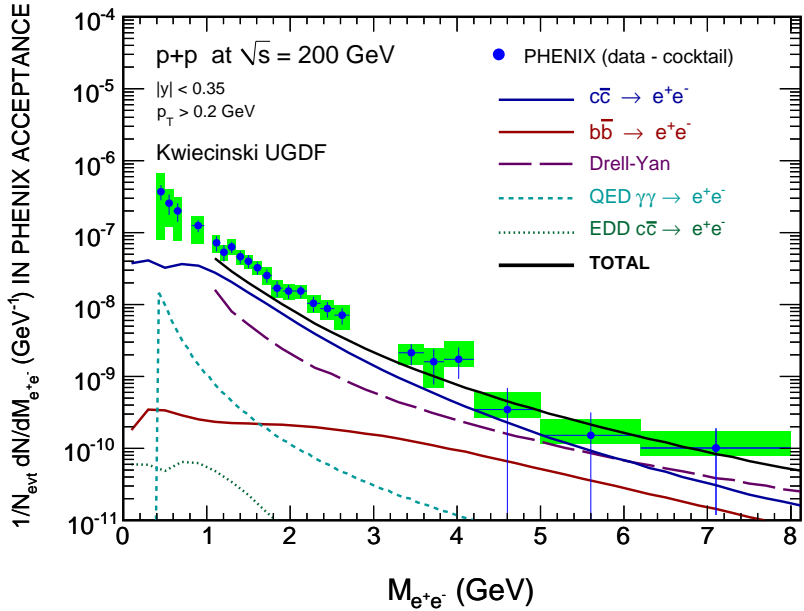}}
\end{minipage}
   \caption{
 \small Dielectron invariant mass distribution for proton-proton
collisions at $\sqrt{s}$ = 200 GeV. Different contributions are
shown separately: semileptonic decay of charm by the upper solid line (blue online), 
semileptonic decay of bottom by the lower solid line (red online), Drell-Yan mechanism by
the long-dashed line, gamma-gamma processes by the short-dashed line (blue online) and 
the EDD contribution by the dotted line (green online).
In this calculation we have included azimuthal angle acceptance of the
PHENIX detector \cite{PHENIX}.
}
 \label{fig:sources-kwiec}
\end{figure} 
%-----------------------------------------------------------------------------

In Fig.\ref{fig:sources-kmr} and Fig.\ref{fig:sources-kutak} we present the same distributions
but here the calculation of heavy quarks is performed 
with the KMR and Kutak-Stasto UGDFs, respectively. One can see that the KMR UGDF gives quite good description of the PHENIX
data in the whole considered dielectron invariant mass range.
Similar results have been obtained with the Kwieci\'nski UGDF except 
very low dielectron invariant masses.
%--------------------------------------------------------------------------
\begin{figure}[!h]
\begin{minipage}{0.75\textwidth}
 \centerline{\includegraphics[width=1.0\textwidth]{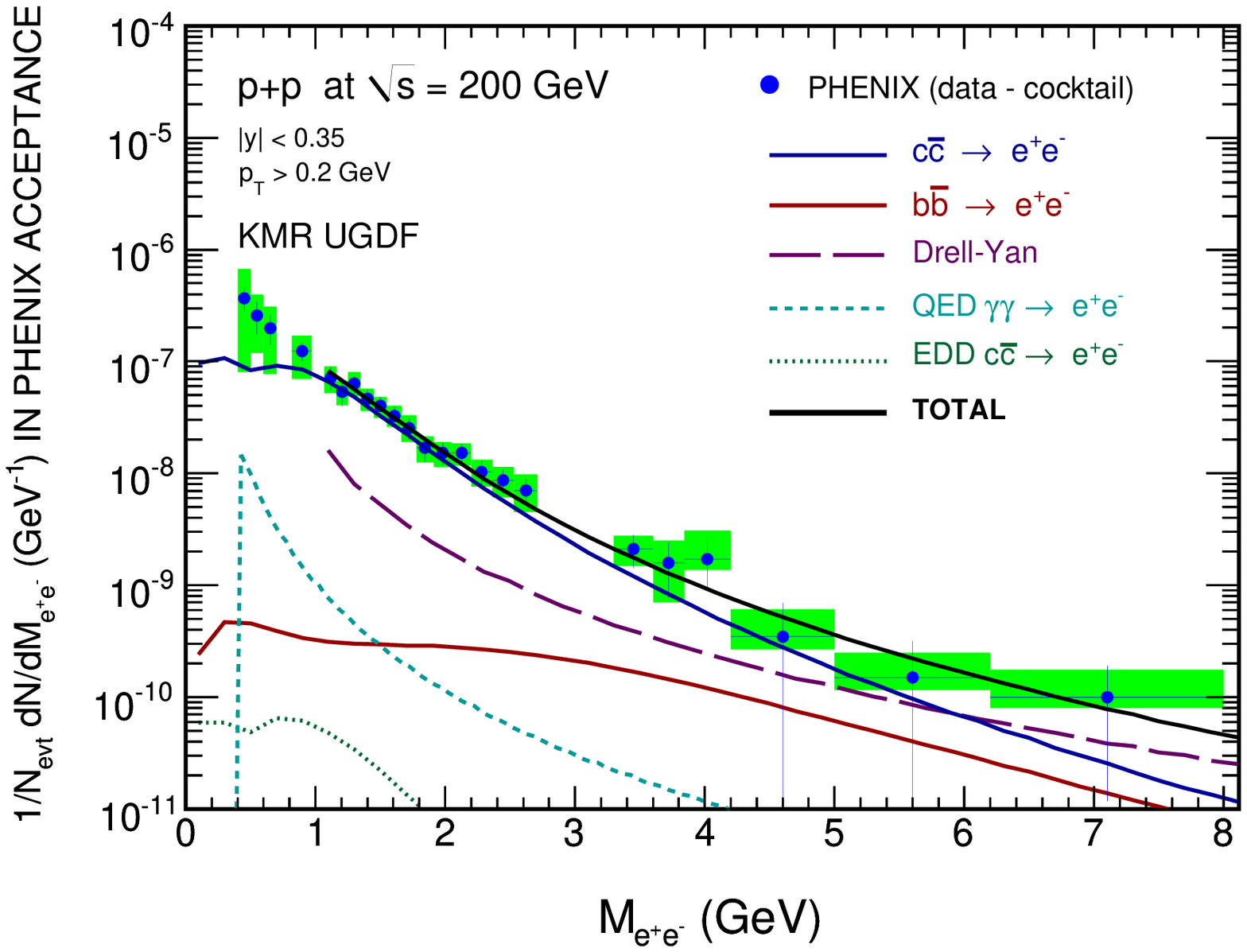}}
\end{minipage}
   \caption{
 \small The same as in Fig.\ref{fig:sources-kwiec} but open charm and bottom components are calculated using
 KMR UGDFs.
}
 \label{fig:sources-kmr}
\end{figure} 
%-----------------------------------------------------------------------------
%--------------------------------------------------------------------------
\begin{figure}[!h]
\begin{minipage}{0.75\textwidth}
 \centerline{\includegraphics[width=1.0\textwidth]{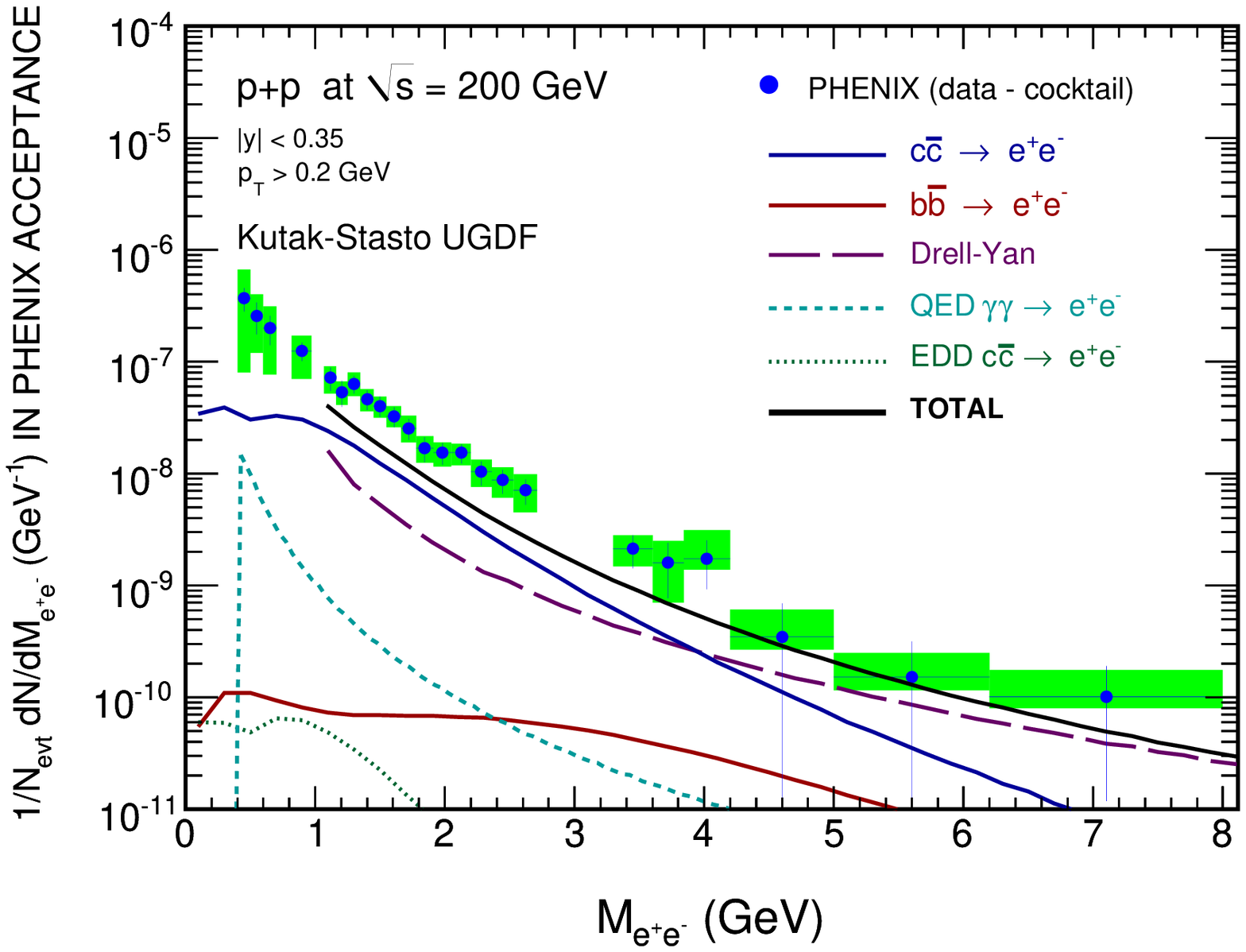}}
\end{minipage}
   \caption{
 \small The same as in Fig.\ref{fig:sources-kwiec} but open charm and bottom components are calculated using
 Kutak-Stasto UGDFs.
}
 \label{fig:sources-kutak}
\end{figure} 
%-----------------------------------------------------------------------------
 Figure \ref{fig:sources-kutak} shows that the Kutak-Stasto UGDFs reproduce the 
experimental data only at large dielectron invariant masses but that is the region where Drell-Yan mechanism
gives significant contribution. 
As was mentioned by the authors of \cite{Kutak}, the Kutak-Stasto UGDF is dedicated exclusively to small-x processes ($x < 10^{-2}$),
so its use for RHIC is at the border of its applicability,
especially for bottom quarks.} 
 In the calculation of heavy quark/antiquark we
have taken $m_c$ = 1.5 and $m_b$ = 4.75 GeV, rather conservative values.\footnote{Often
too small values of heavy quark masses are taken in the calculation
to describe the data.}

\begin{figure}[!h]
\begin{minipage}{0.55\textwidth}
 \centerline{\includegraphics[width=1.0\textwidth]{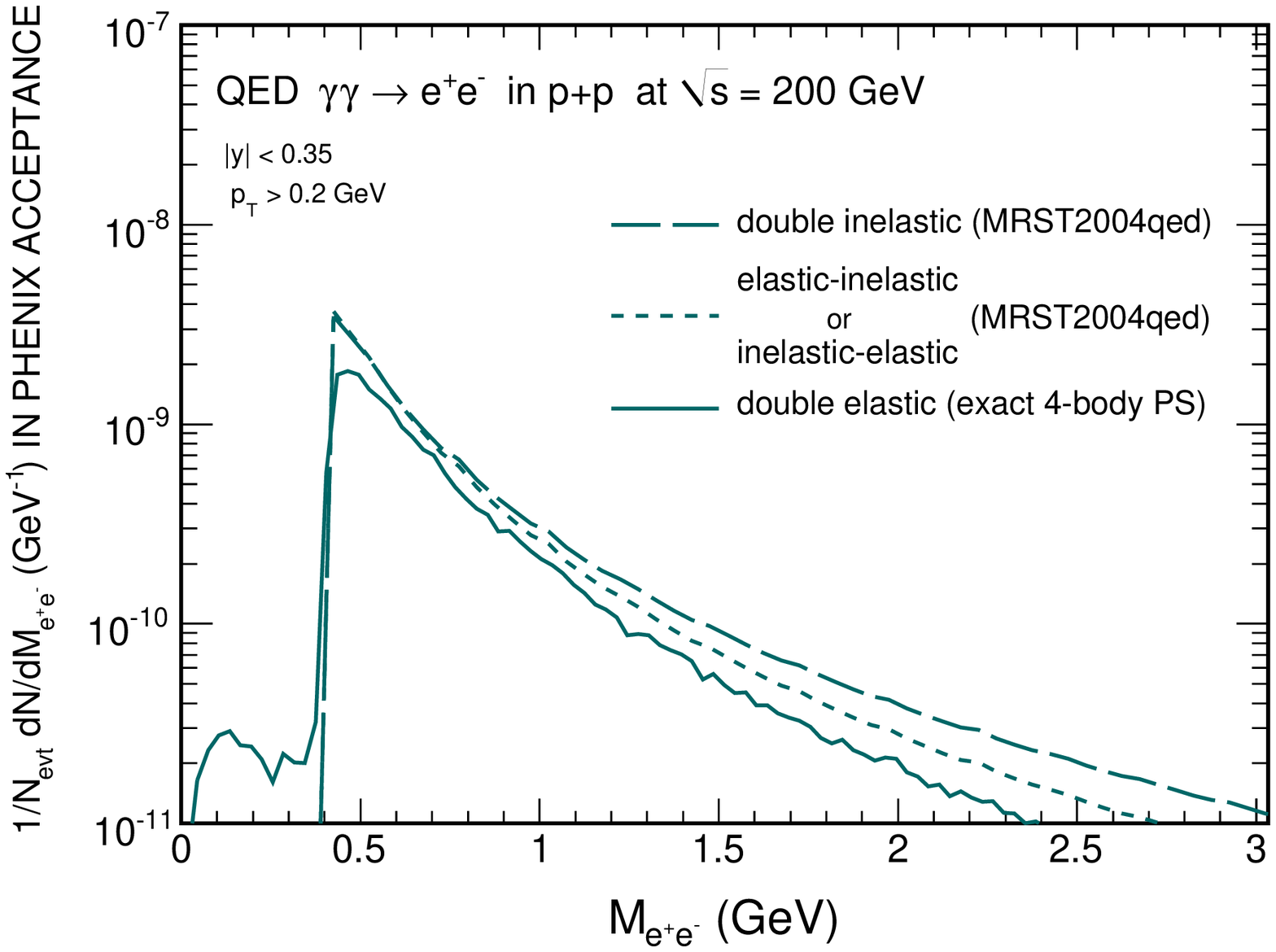}}
\end{minipage}
   \caption{ Contributions of photon-induced mechanisms to
the dielectron invariant mass distributions. The PHENIX detector
limitations are given in the upper-left corner.
}
 \label{fig:gamma_sources}
\end{figure}

In Fig.\ref{fig:gamma_sources} we show separately contributions of
different photon induced mechanisms shown in
Fig.\ref{fig:qed}. The amplitude of the double elastic
contribution is calculated as explained in subsection \ref{sec:qed}, 
the other contributions are calculated in the collinear approximation 
as explained in the same subsection.

%--------------------------------------------------------------------
%----------------------------------------------------------------------

\begin{figure}[!h]
\begin{minipage}{0.47\textwidth}
 \centerline{\includegraphics[width=1.0\textwidth]{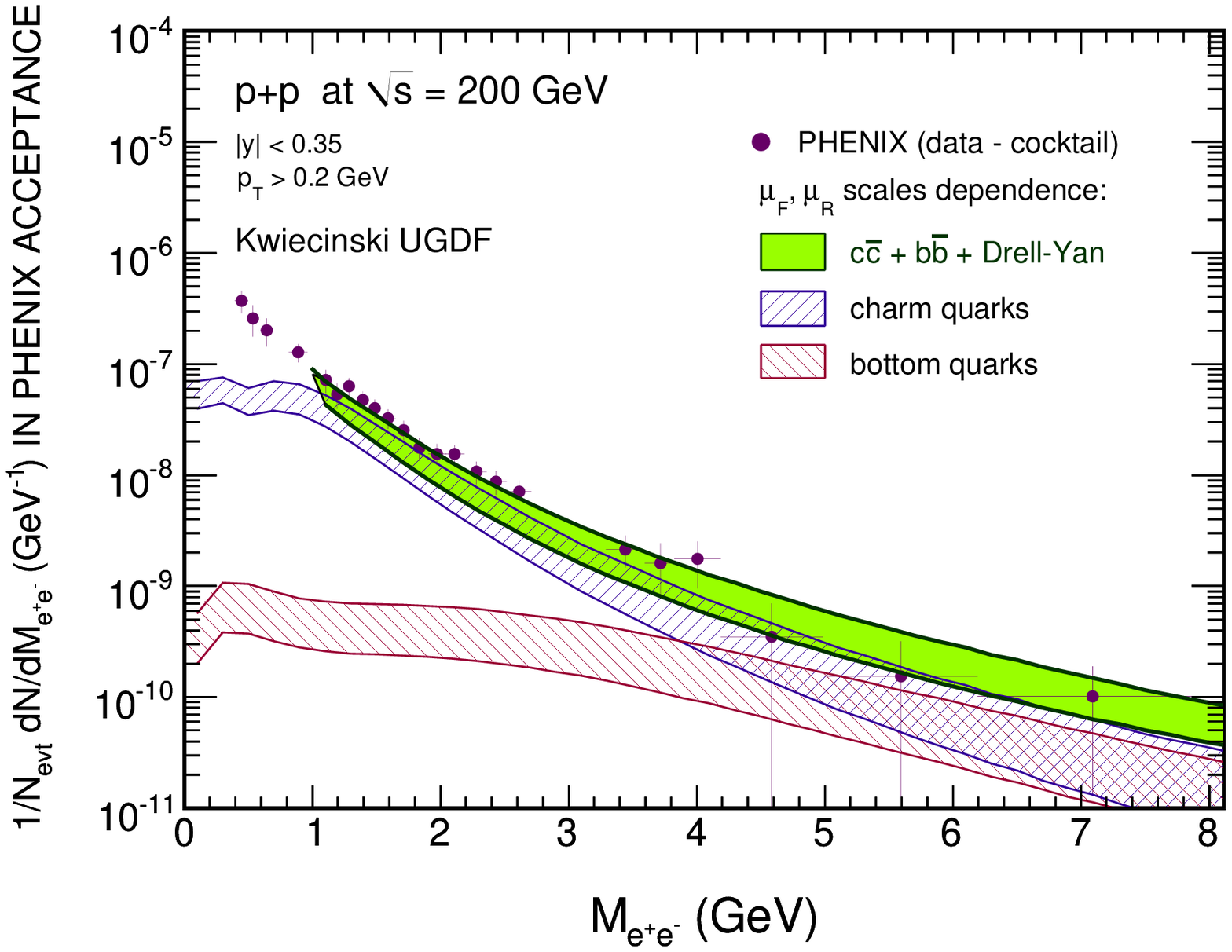}}
\end{minipage}
\hspace{0.5cm}
\begin{minipage}{0.47\textwidth}
 \centerline{\includegraphics[width=1.0\textwidth]{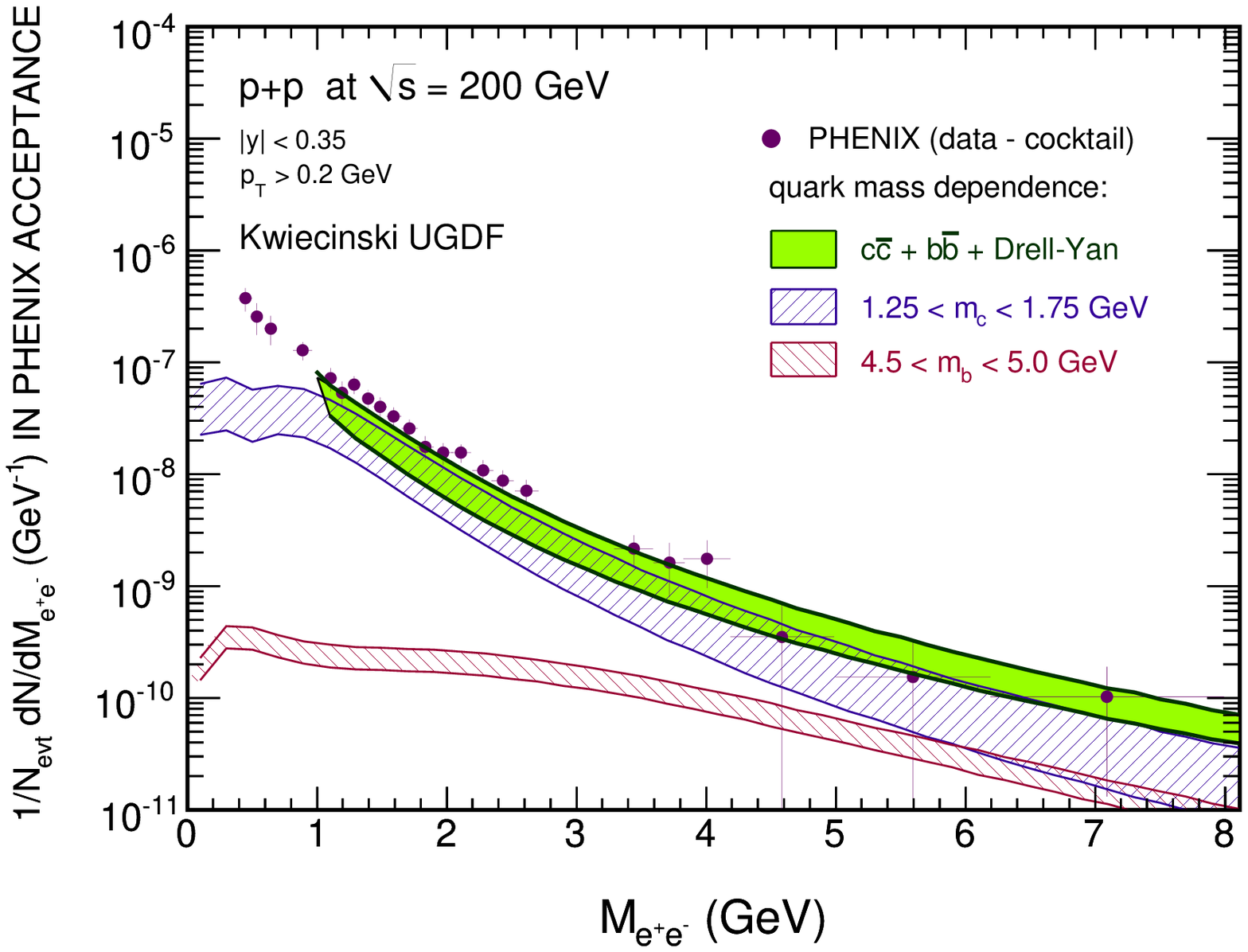}}
\end{minipage}
   \caption{
 \small The uncertainties of theoretical calculations.
The left panel shows the factorization scale uncertainties, 
the lower curve corresponds to $\mu_F^2, \mu_R^2 = m_{1,t}^2 + m_{2,t}^2$ and 
the upper curve to $\mu_R^2 = k_t^2$, $\mu_F^2 = 4 m_Q^2$, where $k_t$ 
is gluon transverse momentum.
The right panel shows the quark mass uncertainties as indicated
in the figure.
}
 \label{fig:uncertainties}
\end{figure}

\begin{figure}[!h]
\begin{minipage}{0.75\textwidth}
 \centerline{\includegraphics[width=1.0\textwidth]{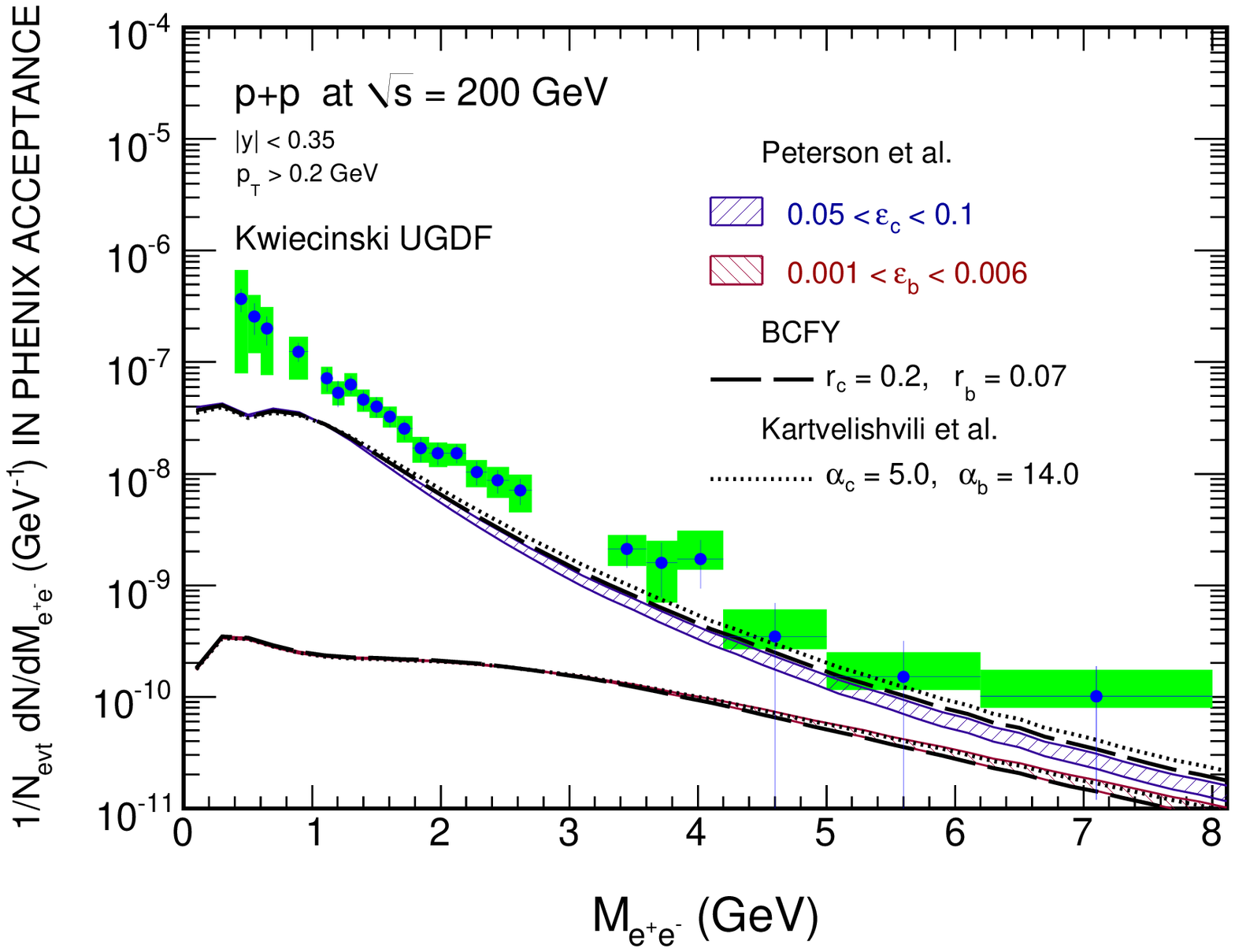}}
\end{minipage}
   \caption{
 \small The uncertainties of theoretical calculations of open charm and bottom related to 
 the choice of the fragmentation functions. BCFY means that fragmentation functions from Ref.\cite{bcfy} were used.
}
 \label{fig:fragmentation}
\end{figure}

In Fig.\ref{fig:uncertainties} we discuss uncertainties related to
the contribution of semileptonic decays. Complementary the left panel 
presents uncertainties due to the factorization scale variation as described
in the figure caption.
The right panel shows uncertainties due to the 
modification of the heavy quark masses ($m_c \in$ (1.25 GeV, 1.75 GeV)
and $m_b \in$ (4.5 GeV, 5 GeV)).
%--------------------------------------------------------------------------
%-----------------------------------------------------------------------------
Figure \ref{fig:fragmentation} presents uncertainties of our predictions
related to the different models of heavy quark fragmentation.
Two bands (blue online and red online for charm and bottom, respectively), estimated with Peterson fragmentation functions show 
sensitivity of our results to the variation of $\epsilon$ parameters in intervals 
specified in the figure. The long-dashed lines represent Braaten et al. perturbative fragmentation model 
and dotted lines are for the fragmentation function proposed by Kartvelishvili et al..
For each of the function we use parameters taken from the literature \cite{PDG,Bracinik}.
As one can observe, in comparison with uncertainties discussed before there is only a small sensitivity of the results 
to the fragmentation functions. Some small differences start to appear only at large dilepton invariant masses where
the error bars of the experimental data are realy large. Besides, in the case of bottom quarks 
such effects are almost negligible. 

%--------------------------------------------------------------------------
%-----------------------------------------------------------------------------

\begin{figure}[!h]
\begin{minipage}{0.325\textwidth}
 \centerline{\includegraphics[width=1.0\textwidth]{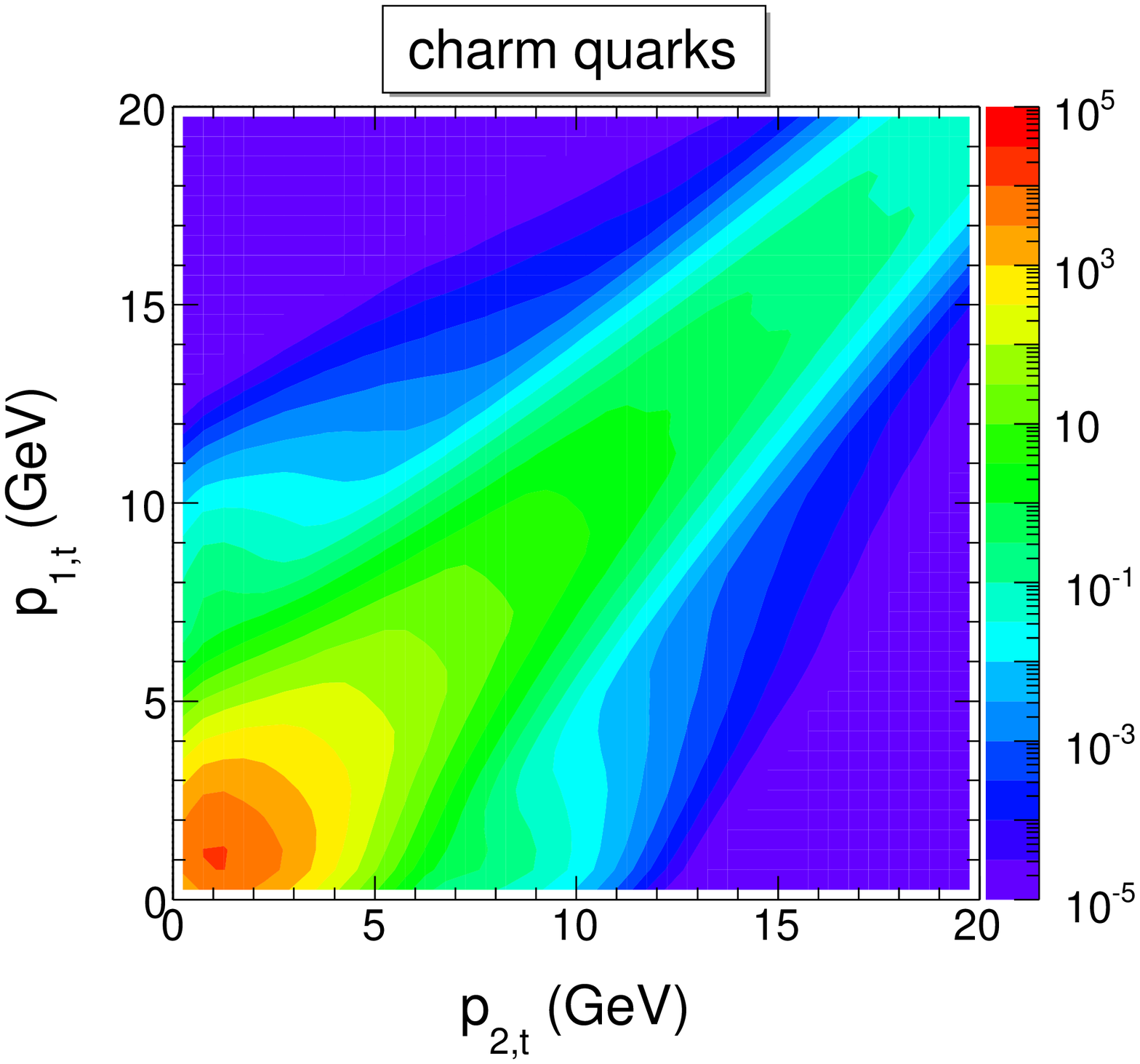}}
\end{minipage}
%\hspace{0.5cm}
\begin{minipage}{0.325\textwidth}
 \centerline{\includegraphics[width=1.0\textwidth]{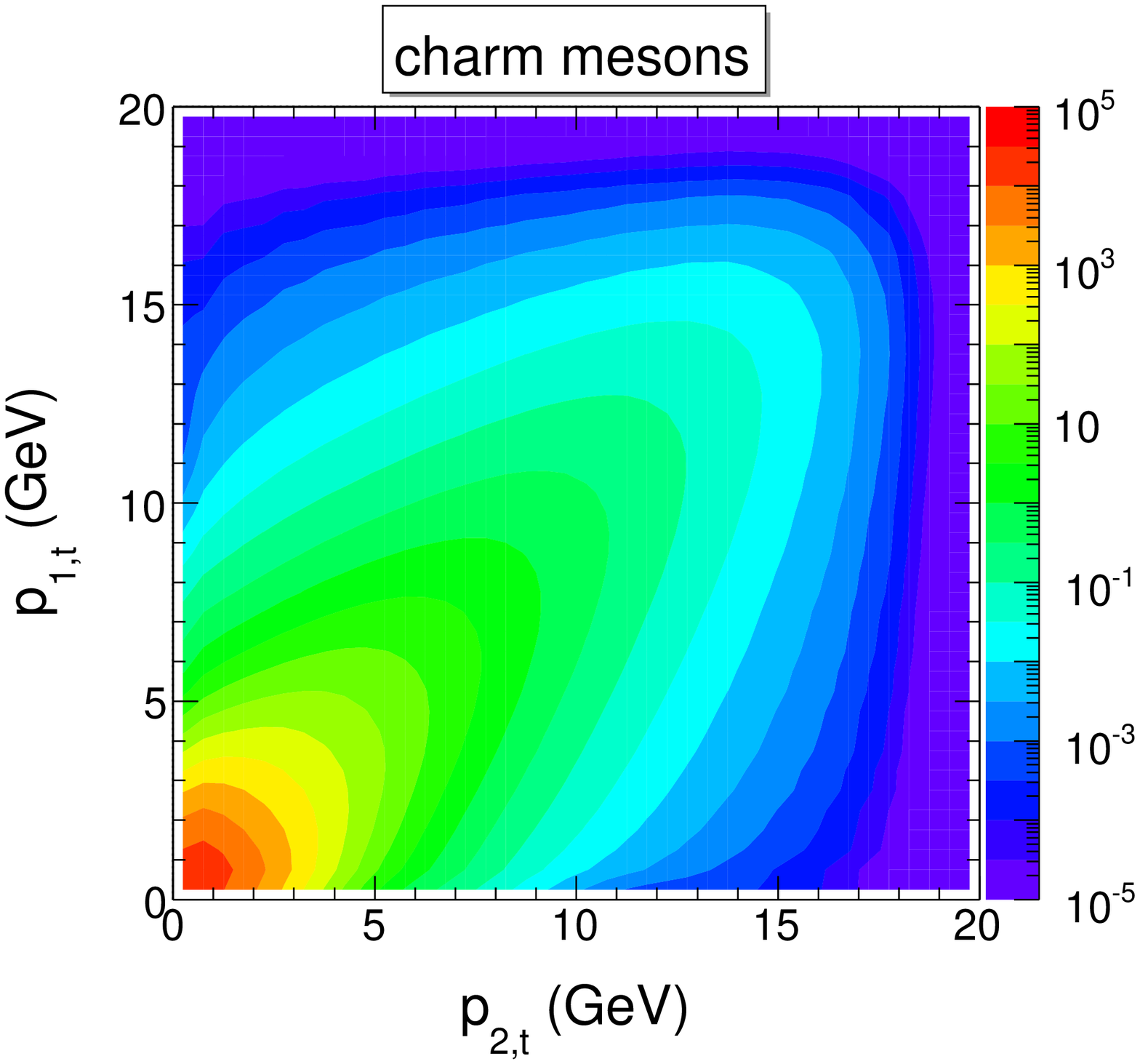}}
\end{minipage}
%\hspace{0.5cm}
\begin{minipage}{0.325\textwidth}
 \centerline{\includegraphics[width=1.0\textwidth]{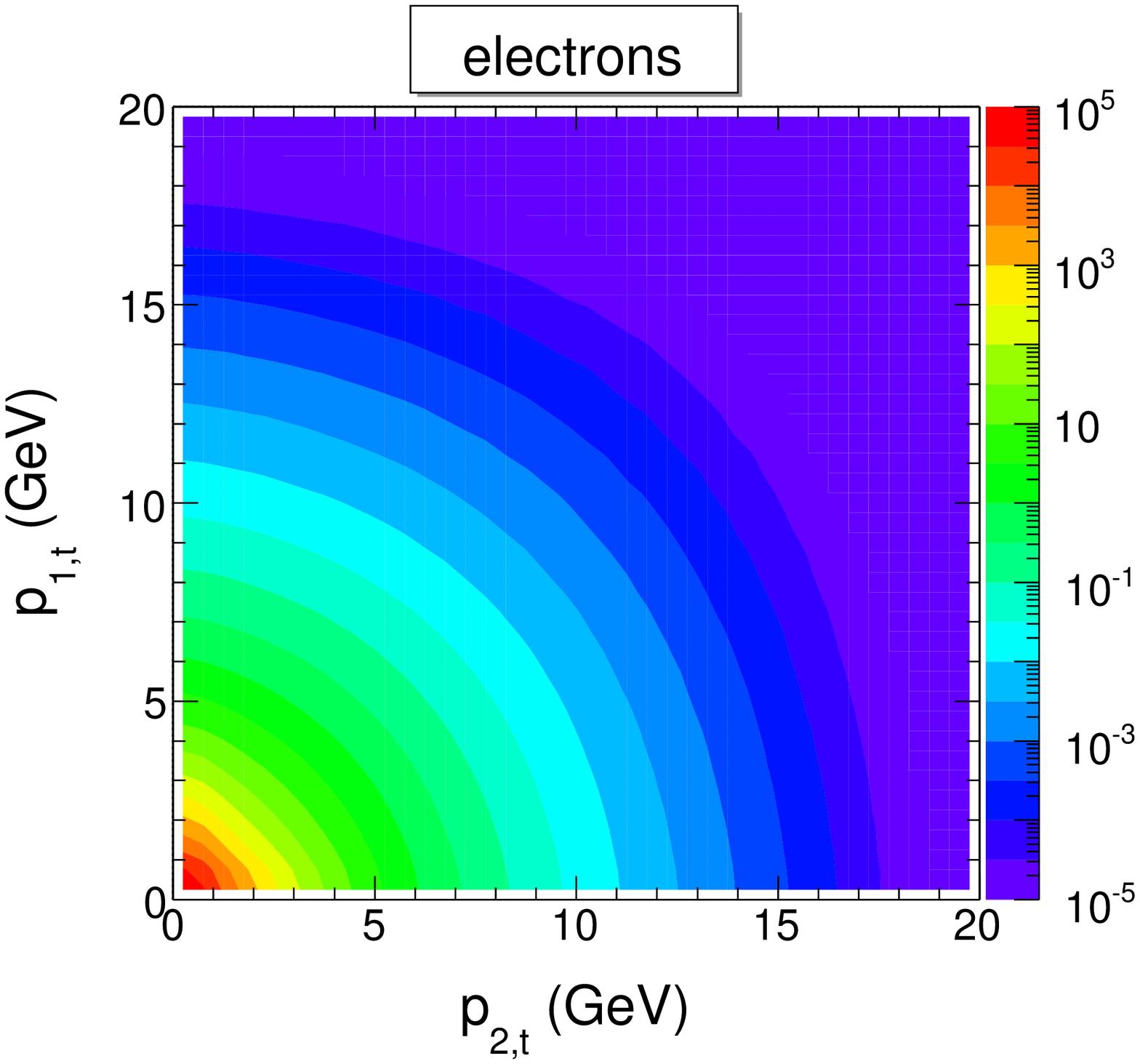}}
\end{minipage}
   \caption{
 \small Two-dimensional distribution in transverse momenta of
$c \bar c$ (left panel), $D \bar D$ (middle panel) and $e^+ e^-$ 
(right panel). Here Kwiecinski UGDF and Peterson fragmentation function were
used.
}
 \label{fig:p1t_vs_p2t}
\end{figure}

Since the transverse momenta of electrons can be measured, one can look 
not only at their distributions but also at correlations between them. 
In Fig.\ref{fig:p1t_vs_p2t} we show two-dimensional distribution 
in transverse momenta of $c$ and $\bar c$ (left panel), 
$D$ and $\bar D$ mesons (middle panel) and $e^+$ and $e^-$ (right panel). 
In contrast to leading-order collinear QCD calculations
already the distribution at the parton level is dispersed
along diagonal. It is further broadened by (assumed independent) fragmentation
process and even more by (assumed independent) semileptonic decays.
So the intial transverse momentum correlations of $c$ and $\bar c$ 
are practically lost when going to the electrons/positrons but 
are interesting and provide a new possibility to test the dynamics of 
the process and our understanding of QCD at work.

%--------------------------------------------------------------------------
%-----------------------------------------------------------------------------

\begin{figure}[!h]
\begin{minipage}{0.55\textwidth}
 \centerline{\includegraphics[width=1.0\textwidth]{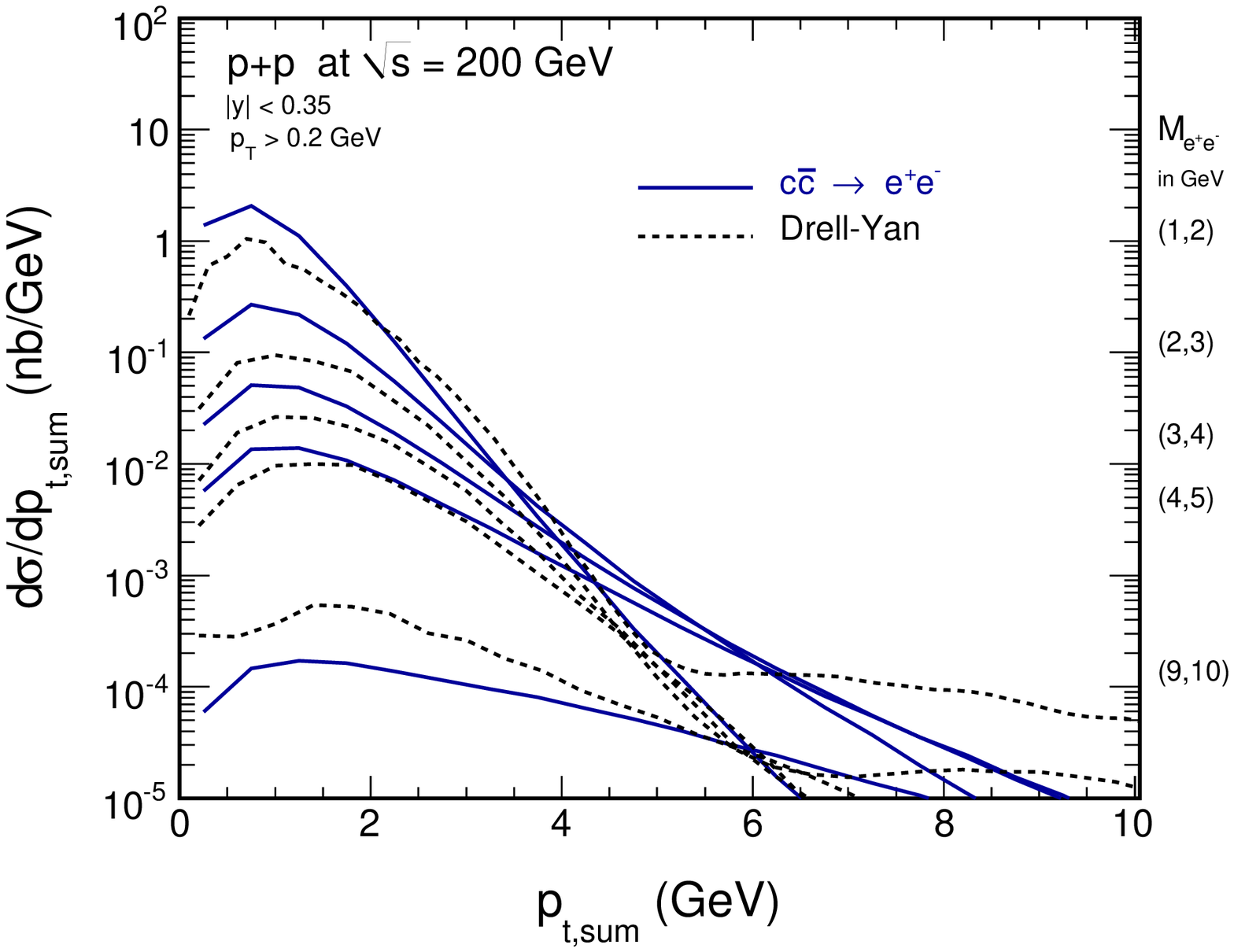}}
\end{minipage}

   \caption{
 \small
Distribution in transverse momentum of the dielectron pair 
for semileptonic decays (solid line) 
and Drell-Yan processes (dashed line). Here Kwiecinski UGDF and Peterson fragmentation function were
used.
}
 \label{fig:ptsum}
\end{figure}

If the detector can measure both transverse momenta of an electron/positron
and its directions, as the STAR detector at RHIC can do,
one can construct a distribution in transverse momentum of the dielectron
pair: $\vec{p}_{t,sum} = \vec{p}_{1t} + \vec{p}_{2t}$.
Our predictions for the semileptonic decays and Drell-Yan processes
are shown in Fig.\ref{fig:ptsum}. Both processes give rather 
similar distributions. To our knowledge the distributions 
of this type were never measured experimentally as they cannot 
easily be compared to the calculations in the collinear approach due 
to its inherent singularities.
Obviously this is not the case for the $k_t$-factorization approach
discussed in the present analysis.
The distribution in $p_{t,sum}$ is not only a consequence
of gluon transverse momenta, as it is for quark and antiquark production,
but invlolves also fragmentation process and semileptonic decays. 
A measurement of this quantity would test then all stages of the process.

%--------------------------------------------------------------------------
%\vspace{-2mm}
\begin{figure}[!h]
\begin{minipage}{0.55\textwidth}
 \centerline{\includegraphics[width=1.0\textwidth]{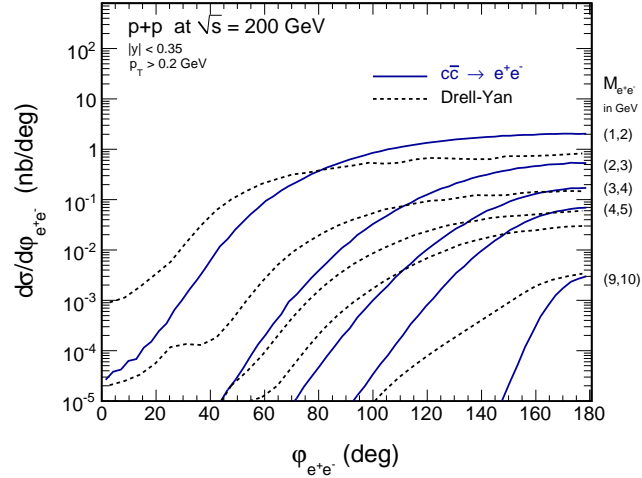}}
\end{minipage}
   \caption{
 \small
Distribution in azimuthal angle between electron and positron 
for semileptonic decays (solid line) 
and Drell-Yan processes (dashed line). Here Kwiecinski UGDFs and Peterson fragmentation functions were
used.
}
 \label{fig:azimuth}
\end{figure}

%----------------------------------------------------------------------------
With good azimuthal granulation of detectors one could construct
distribution in azimuthal angle between electron and positron. 
Our corresponding predictions are shown in Fig.\ref{fig:azimuth}. One can see 
an interesting dependence on the invariant mass of the dielectron pair
-- the smaller the invariant mass the large the decorrelation in azimuthal
angle.
%--------------------------------------------------------------------------

%-----------------------------------------------------------------------------

%-----------------------------------
\section{Conclusions}
%-----------------------------------

In the present analysis we have discussed correlations of charmed mesons
and dielectrons at the energy of recent RHIC experiments. We have calculated the spectra
in dielectron invariant mass, in azimuthal angle between electron and positron as
well as for the distribution in transverse momentum of the pair.
The uncertainties due to the choice of UGDFs, choice of the factorization and renormalization scales,
choice of the heavy quark masses as well as fragmentation functions
have been quantified. The uncertainties for UGDFs are larger than those
for fragmentation functions.
We have obtained good description of the dielectron invariant mass
distribution measured recently by the PHENIX collaboration at RHIC. 

The contribution of electrons from Drell-Yan processes is only slightly 
smaller than that from the semileptonic decays. The distributions in
azimuthal angle between electron and positron and in the transverse 
momentum of the dielectron pair from both processes are rather similar. 
We do not find a possibility of a clear separation of both processes.
It was found that the distribution in azimuthal angle strongly depends
on dielectron invariant mass.

We have also included exclusive double-diffractive contribution
discussed recently in the literature.
At the rather low RHIC energy it gives, however, a very small contribution
to the cross section and can be safely ignored. It may not be the case 
at the LHC energy, as the EDD contribution grows much faster than 
the inclusive cross section.

The QED double-elastic, double-inelastic, elastic-inelastic and 
inelastic-elastic processes give individually
rather small contribution but when added together are not negligible
especially at low dielectron invariant masses where some strength is 
clearly missing \cite{PHENIX}.

In the present analysis we have studied correlations between electron
and positron and in some cases between mesons. It can be also
interesting to look at correlations between a $D$ meson and electron.
This will be a subject of a forthcoming analysis.

\vspace{20mm}

{\bf Acknowledgments}\\
We are indebted to Andre Mischke for exchange of 
information on recent RHIC results and very useful 
comments. This work was partially supported by the Polish
grants MNiSW No. N202 2492235 and MNiSW No. N202 236937.

%=========================================================================================

\end{document}